\documentclass[10pt,conference]{IEEEtran}
\IEEEoverridecommandlockouts

\usepackage{cite}
\usepackage{url}
\usepackage{bbm}
\usepackage{amsmath,amsfonts} 
\usepackage{float}
\usepackage{graphicx}
\usepackage{algorithm}
\usepackage{algpseudocode}
\usepackage{siunitx}
\usepackage[labelfont=bf]{caption}
\usepackage{subcaption}
\usepackage{tikz}
\usepackage{pgfplots}
\usepackage{epstopdf}
\usepackage{multirow}
\usepackage[font=scriptsize]{caption}
\usepackage{soul}
\usepackage{color, colortbl}
\usepackage{array}
\usepackage[T1]{fontenc}
\usepackage[utf8]{inputenc}
\usepackage{mathptmx}
\usepackage{listings}
\usepackage{adjustbox}
\usepackage{makecell}

\usepackage{enumitem}
\usepackage{hyperref}
\usepackage{pifont}

\def\BibTeX{{\rm B\kern-.05em{\sc i\kern-.025em b}\kern-.08em
    T\kern-.1667em\lower.7ex\hbox{E}\kern-.125emX}}

\begin{document}

\title{RAPTOR: Advanced Persistent Threat Detection in Industrial IoT via Attack Stage Correlation}


\author{\IEEEauthorblockN{Ayush Kumar}
\IEEEauthorblockA{\textit{Cyber Security Strategic Technology Centre} \\
\textit{ST Engineering} \\
Singapore, Singapore \\
ayush.kumar@u.nus.edu}
\and
\IEEEauthorblockN{Vrizlynn L.L. Thing}
\IEEEauthorblockA{\textit{Cyber Security Strategic Technology Centre} \\
\textit{ST Engineering} \\
Singapore, Singapore \\
vriz@ieee.org}
}
			


\maketitle

\begin{abstract}
Past Advanced Persistent Threat (APT) attacks on Industrial Internet-of-Things (IIoT), such as the 2016 Ukrainian power grid attack and the 2017 Saudi petrochemical plant attack, have shown the disruptive effects of APT campaigns while new IIoT malware continue to be developed by APT groups. Existing APT detection systems have been designed using cyberattack TTPs modelled for enterprise IT networks and leverage specific data sources (e.g., Linux audit logs, Windows event logs) which are not found on ICS devices. In this work, we propose \textsf{RAPTOR}, a system to detect APT campaigns in IIoT. Using cyberattack TTPs modelled for ICS/OT environments and focusing on `invariant' attack phases, \textsf{RAPTOR} detects and correlates various APT attack stages in IIoT leveraging data which can be readily collected from ICS devices/networks (packet traffic traces, IDS alerts). Subsequently, it constructs a high-level APT campaign graph which can be used by cybersecurity analysts towards attack analysis and mitigation. A performance evaluation of \textsf{RAPTOR}'s APT attack-stage detection modules shows high precision and low false positive/negative rates. We also show that \textsf{RAPTOR} is able to construct the APT campaign graph for APT attacks (modelled after real-world attacks on ICS/OT infrastructure) executed on our IIoT testbed. 
\end{abstract}

\begin{IEEEkeywords}
Industrial Internet of Things, IoT, IIoT, Advanced Persistent Threat, APT, APT Detection
\end{IEEEkeywords}

\section{Introduction}
\label{intro}
The Internet of Things (IoT) is a network of sensing devices with low-power and limited processing capability, which exchange data with each other and/or systems (e.g., gateways, cloud servers), normally using wired and wireless technologies. Industrial IoT (IIoT) refers to the extension of IoT in industrial sectors and applications. With a strong focus on machine-to-machine (M2M) communication, big data, and machine learning, the IIoT enables industries and enterprises to have better efficiency and reliability in their operations. What makes IIoT distinct from IoT is the intersection of information technology (IT) and operational technology (OT). However, this convergence has widened the attack surface and increased the potential risks of cyberattacks being launched against such critical systems. A more significant concern relates to legacy OT systems (i.e., brownfield IIoTs) which are usually isolated but are becoming more connected with new IT technologies. Sophisticated attackers such as those belonging to Advanced Persistent Threat (APT) groups can easily gain access to such brownfield IIoT systems and damage their operation for lengthy periods of time.

An APT is a type of cyberattack that is designed to gain unauthorized access to a computer network and remain undetected for an extended period of time. APTs are usually carried out by skilled hackers who use sophisticated techniques including custom malware to infiltrate and evade security measures. These attacks can have serious consequences for large enterprises or governmental networks, such as intellectual property theft, compromised personal data, or critical infrastructure damage. APT attacks are often carried out by nation-state actors or state-sponsored groups, but they can also be conducted by non-state actors. Some of the most well-known APT groups include APT10, APT28, and APT41.


IIoT systems are more prone to attacks by APT adversaries than traditional Industrial Control Systems (ICS)/Operational Technology (OT) networks \cite{checkpoint-iiot-apt} mainly due to addition of connectivity to IT networks (enabling lateral movement for attackers). Furthermore, ICS assets themselves are prime targets for APT campaigns. This is because ICS devices often run on legacy, proprietary software which were not designed with security in mind and are not patched/updated regularly due to concerns over downtime in critical systems. APT attacks can be used to gather ICS-related intelligence, disrupt industrial processes, shut down critical systems and endanger human lives. Such APT attacks by well-resourced groups have happened a number of times in the past, e.g., 
the 2016 attack on Ukranian transmission level substation which cut a fifth of the capital Kyiv off power for an hour \cite{ukraine-2016} and the 2017 attack on a Saudi petrochemical plant which almost shut down the plant's safety controllers which could have caused an explosion \cite{saudi-2017}. \textit{Industroyer/Crashoverride}, the malware behind 2016 Ukraine power grid attack and \textit{TRITON}, the malware behind 2017 Saudi petrochemical plant attack are still active, targeting electrical substations and energy utilities respectively \cite{industroyer2, triton2}. New IIoT malware such as \textit{Incontroller/Pipedream} \cite{incontroller}, which was revealed as recently as 2022, contains modules that target specific ICS devices such as OPC servers, Schneider Electric PLCs using Modbus and Codesys protocols, and Omron PLCs and servo drives.    

Security solutions which are commonly deployed in IT networks such as firewalls, NIDS (Network Intrusion Detection System), SIEM (Security Information and Event Management) products are not common in ICS/OT networks. Even if they are deployed, either they cannot correlate low-level alerts into a high-level representation of an ongoing APT campaign or they cannot precisely map the correlated alerts to actual APT attack steps taking place on different hosts/devices over a long period of time. Thus, there is an impending need to design systems which can precisely detect ongoing APT attack campaigns in IIoT environments early before they cause substantial damage. 

An APT campaign typically consists of various stages which occur one after another (this will be explained later in the paper). However, not all stages defined in existing attack frameworks are found together in real-world APT campaigns. Each stage in an APT campaign is linked to the previously executed stages, e.g., in terms of chronology of execution, target hosts affected. Thus, if we are able to detect some of the individual attack stages, the above fact can be exploited to correlate the detected stages and reconstruct the APT campaign with some acceptable margin of error.
In this work, we present \textsf{RAPTOR}, a system for detecting ongoing APT campaigns in IIoT environments. It uses cyberattack  Tactics, Techniques and Procedures (TTPs) modelled for ICS/OT environments, removing unnecessary attack phases and focusing on the `invariant' ones. \textsf{RAPTOR} consists of an \textit{APT attack-stage detection and correlation engine} which takes as input a variety of readily available, non-proprietary data sources in the context of ICS devices (network traffic traces, IDS alerts) and detects attack stages which are part of an ongoing APT campaign. The attack stages are then stitched together based on their attributes to produce the APT campaign graph, which is a high-level representation of APT activity across the target IIoT network. As shown in Section \ref{results}, a performance evaluation of \textsf{RAPTOR}'s APT attack-stage detection modules reveals high precision and low false positive/negative rates.

The main contributions of our work are as follows:
\begin{itemize}
	\item We present, \textsf{RAPTOR}, a system for detecting ongoing APT campaigns in IIoT environments which focuses on certain `invariant' APT attack phases.
	\item We employ an attack-stage detection and correlation approach which uses readily available, non-proprietary ICS data sources for its operation.
	\item \textsf{RAPTOR} constructs a compact, high level APT campaign graph which can be useful for cybersecurity analysts.
	\item We evaluate \textsf{RAPTOR}'s performance using a new dataset which includes attack TTPs close to real-world APT attacks on ICS/OT environments. 
\end{itemize}
 

\section{Related Work}
\label{apt-det-literature}
APT detection in enterprise network settings has received significant attention in computer security literature in recent years. Milajerdi et al. \cite{holmes} have presented \textit{HOLMES}, a system for APT campaign detection with high confidence. It maps activities in host audit logs and enterprise security alerts to the cyber kill-chain, correlates the alerts generated by APT steps based on information flow between low-level entities (files, processes, etc.) and builds a high-level scenario graph encapsulating the attack TTPs and information flows between entities involved in the TTPs. \textit{CONAN} \cite{conan} is an APT detection system which makes two major modifications to HOLMES's approach: one, it utilizes a state-based detection framework where all processes and files are represented as data structures similar to finite-state automata and two, it focuses on three constant attack phases of APTs. Wilkens et al. \cite{kcsm} have proposed a method to construct APT attack graphs from IDS (Intrusion Detection System) alerts to assist human analysts. IDS alerts are clustered into meta-alerts and single alerts, assigned potential attack stages and finally used to synthesize APT scenario graphs based on a \textit{kill chain state machine}. 

Irshad et al. \cite{trace} have proposed \textit{TRACE}, a provenance tracking system for enterprise-wide APT detection. TRACE offers host-level provenance tracking at the granularity of program executions units and integrates provenance collected from individual hosts to construct distributed enterprise-wide causal graphs. Hopper \cite{hopper} instead focuses on the detection of a single APT attack stage: lateral movement, within enterprise networks. It does so by building a graph of login activity among machines in a network and identifying suspicious login sequences. A path inference algorithm is then deployed to identify the broader paths that each login belongs to, "caused" by the same user. Finally, an anomaly detection algorithm is applied to conservatively infer the set of login paths most likely to reflect lateral movement. Han et al. \cite{unicorn} have presented \textit{UNICORN}, which uses anomaly detection based on system data provenance to detect APT campaigns. UNICORN ingests a streaming whole-system provenance graph captured on one or more networked hosts, computes a histogram over the graph, transforms the graph to a fixed-size incrementally updatable graph sketch and finally clusters the sketches to generate a system-wide execution model (during training) to which incoming sketches during deployment are compared. Recently, Mahmoud et al. \cite{apthunter} have proposed \textit{APTHunter} which builds a whole-system provenance graph from OS kernel audit logs and then applies attack provenance queries to the graph to generate alerts for individual APT attack stages, thus enabling it to detect APT campaigns in early stages. The provenance queries are formed by combining MITRE ATT\&CK framework attack TTPs with real-world cyberattack artifacts from Cyber Threat Intelligence reports.  

However, all the above works suffer from certain limitations:
\begin{itemize}
\item \cite{holmes, conan, kcsm, trace, hopper, unicorn, apthunter} focus on enterprise networks only.
\item \cite{holmes, conan, trace, apthunter} have been designed using audit logs-based provenance, \cite{unicorn} depends on Linux Security Modules framework for provenance, \cite{hopper} has been designed using enterprise logs, and \cite{kcsm} employs network IDS alerts. Since each of these works is based on a single data source, they are unable to leverage the information provided by other data sources.
\item Exclusively graph-based approaches \cite{holmes, trace, apthunter} tend to suffer from dependency explosion problem which means that with time, each graph node gives rise to many edges which in turn give rise to new nodes and so on. A backward trace through the graph to infer a path exponentially increases the number of probable paths.
\item Except for \cite{kcsm}, no other work mentioned above detects an APT campaign from a whole-network perspective. 
\end{itemize}

Unfortunately, there has been no research work yet on detecting APT campaigns in IIoT settings which is the focus of this paper. IIoT environments differ from enterprise IT infrastructure in that they combine both IT and OT \cite{cisco-it-ot}. 
Therefore, attack tactics, techniques and procedures (TTPs) for IIoT include TTPs for enterprise IT (e.g., reconnaissance, initial access, persistence, privilege escalation) as well as the TTPs for OT (inhibit response function, impair process control, impact). APT detection systems designed for enterprise IT networks cannot be directly applied to IIoT since they do not include OT TTPs and further, they use audit logs collected from hosts to build system-level provenance graphs which is not applicable to ICS devices.

Our work differs from existing works on APT detection in enterprise IT networks \cite{holmes, conan, kcsm, unicorn, apthunter} in the following aspects:
\begin{itemize}
\item The focus is on APT detection in IIoT settings which combines both IT and OT environments instead of just enterprise IT networks. The APT attack stages used to design the detection system in our paper have been adapted to IIoT settings.
\item We leverage data from various sources such as network packet traces, host logs (including audit logs) as well as NIDS/HIDS alerts instead of limiting ourselves to a specific or \textit{proprietary} data source. The optimal data source(s) for detection of each APT attack stage is(are) identified and used.
\item We perform APT campaign detection from a whole-network perspective instead of detecting APT artifacts on individual hosts only. 
\end{itemize}

\section{Background}
\subsection{Existing Attack Frameworks for IIoT}
The authors in \cite{x-iiotid} have proposed a generic IIoT attack life-cycle framework consisting of the following stages: reconnaissance, weaponization, exploitation, lateral movement, command \& control, exfiltration and tampering. MITRE has also released an Adversarial Tactics, Techniques and Common Knowledge (ATT\&CK) framework for ICS \cite{mitre-attack-ics} consisting of the following tactics/stages: initial access, execution, persistence, privilege escalation, evasion, discovery, lateral movement, collection, command and control, inhibit response function, impair process control,  and impact. 

\subsection{IIoT APT Invariant State Machine}
\label{iasm}
It should be noted that real-world APT campaigns in IIoT environments do not follow all the stages in attack frameworks mentioned above. Further, each tactic in an attack framework consists of many techniques, with new ones being added regularly. It is almost impossible to model all the possible attack techniques and then use them towards detection during ongoing APT campaigns. Therefore, using the IIoT attack life-cycle framework \cite{x-iiotid} and the MITRE ATT\&CK framework for ICS \cite{mitre-attack-ics}, we have identified certain attack stages/tactics which are `invariant': \textit{Command-and-control}, \textit{Discovery}, \textit{Lateral movement}, \textit{Fieldbus scanning} and \textit{CE communication spoofing}. These attack tactics consist of only a few techniques and those techniques have not changed significantly across APT campaigns over the years (and are not expected to change significantly in future). It is easier to model these invariant attack tactics and use them towards APT detection. Here, the \textit{Fieldbus scanning} and \textit{CE communication spoofing} attack stages as introduced above are specific to IIoT settings and are not covered in previous works on APT detection. 

The APT attack stages described above might come across as similar to a Red Team Assessment in an organization which is because they are. Red teaming uses various attack tactics, techniques and tools to test an organization's detection and response capabilities with a goal to access target data or systems \cite{rapid7-pentest-redteam}. It has also been used on enterprise testbeds from which data is collected for evaluating the performance of APT detection systems proposed in a few existing works \cite{holmes, trace, conan}.

\begin{figure}[h]
	\centering
	\includegraphics[scale=0.25]{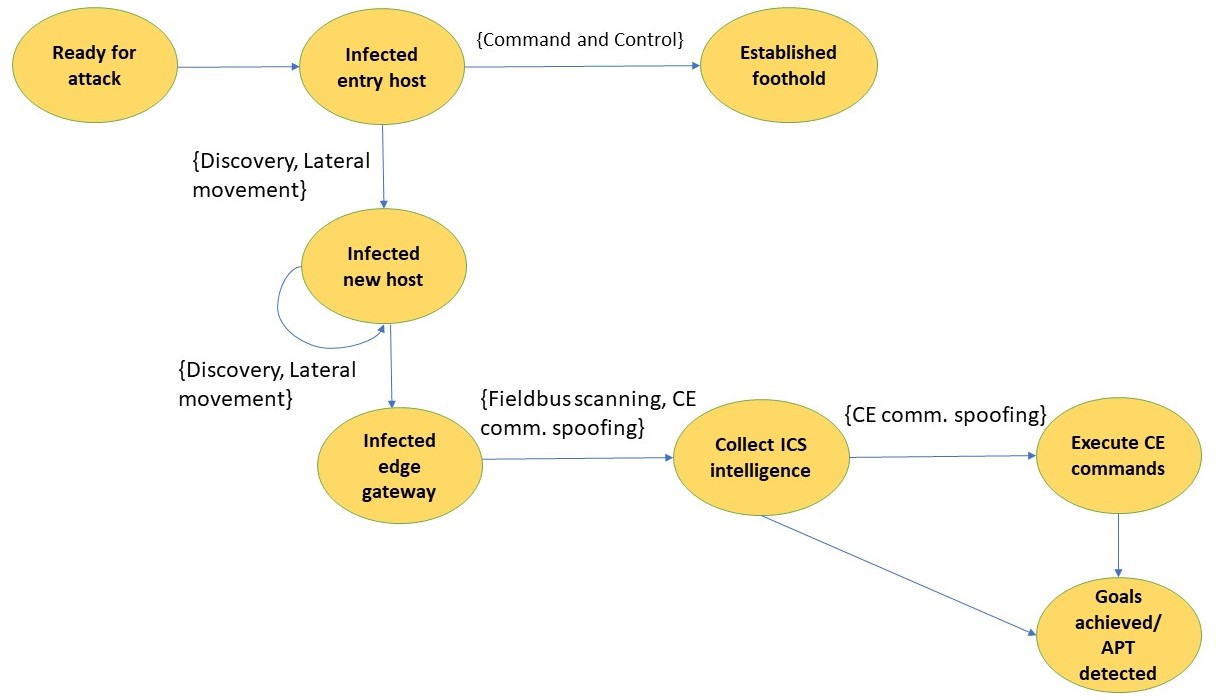}
	\caption{IIoT APT Invariant State Machine}
	\label{apt-inv-state-mach}
\end{figure}

We propose an \textit{IIoT APT Invariant State Machine} (IASM) as shown in Fig. \ref{apt-inv-state-mach} which models a typical APT campaign in an IIoT environment as a finite-state machine. The states in IASM represent the states of the APT campaign while the state transitions are brought about by the deployment of invariant APT tactics identified earlier. Most real-world APT campaigns in IIoT environments such as those described in Section \ref{intro} follow our proposed IASM. 

\textit{IASM Description}: Once the APT attackers have acquired all the resources and information required for attack campaign (\textit{Ready for attack} state), they move by compromising one or more of the public network-facing hosts to gain entry into the target IIoT network (\textit{Infected entry host} state) and establish communication with a Command-and-control or C\&C server (\textit{Establish foothold} state). Next, the attackers scan for other hosts connected to the compromised machine and attempt to gain control of one of the discovered hosts (\textit{Infected new host} state) either by using CVE vulnerabilities or by stealing the remote access credentials for the discovered host and then logging in to it. The attackers may attempt to move across the IIoT network by gaining control of more hosts, thus remaining in the same \textit{Infected new host} state and using the same tactics as outlined earlier. Once the attackers reach the edge gateway (\textit{Infected edge gateway} stage), they scan for control elements (PLC/RTU/SIS) and their slave devices connected via fieldbus protocols. Fieldbus refers to Modbus and other similar open-source or proprietary vendor protocols  (e.g., Profibus/Profinet, CAN) which are used to communicate with respective vendor control elements. Subsequently, they spoof communications with the discovered control elements and their slaves to gain more information about them (\textit{Collect ICS intelligence} state) and execute commands on control elements remotely (\textit{Execute CE commands} state). Depending on the target(s) set by the APT adversaries (just collection of ICS intelligence or execution of desired commands on CE), they might end the campaign wilfully or forcibly due to detection by cybersecurity analysts (\textit{Goals achieved/APT detected} state). 

\subsection{Threat Model} 
We assume that only a few machines in the enterprise tier of the target IIoT network are connected to the Internet (through firewall/IDS). All other machines in the enterprise tier and other tiers (platform, edge) are isolated from the Internet though some of them can still communicate with the Internet-connected machines in enterprise tier. The APT attackers can enter the target IIoT network through the internet-facing enterprise tier machines (remote access) or other machines to which plant operators/engineers have access (insider attack). Once inside the enterprise tier network, the attackers can move laterally across machines till they reach the edge tier consisting of edge gateway, control elements and sensors/actuators. The attackers are assumed to be well-resourced in terms of computational resources, financial backing, hacking skills and time which is true of most APT groups.

\section{System Overview}
\label{system-overview}

\subsection{RAPTOR Architecture}
\begin{figure}[h]
	\centering
	\includegraphics[scale=0.3]{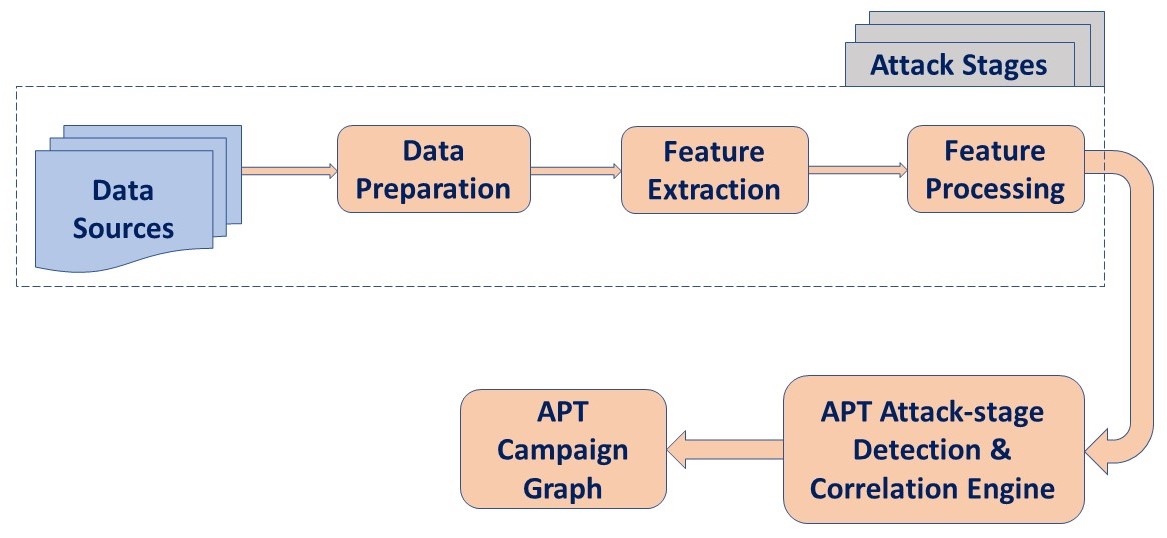}
	\caption{RAPTOR Architecture}
	\label{raptor-arch}
\end{figure}

As shown in Fig. \ref{raptor-arch}, \textsf{RAPTOR} consumes data from different sources such as network traffic traces, \textit{audit} logs, HIDS/NIDS alerts and host logs. The data is processed prior to extraction of features and then the features are processed before being sent to the \textit{APT attack-stage detection \& correlation engine} which detects the invariant APT attack stages in IASM using the optimal data source identified in the following sub-section. The detection methods employed for detection of the attacks-stages are explained in sub-section \ref{apt-attack-det-method}. The attack stages are detected and correlated using their attributes to re-construct the APT campaign which is presented in the form of a graph as described in sub-sections \ref{apt-attack-corr-method} and \ref{apt-campaign-graph}. The APT Campaign Graph (ACG) thus constructed can be utilized by cybersecurity analysts to come up with appropriate actions to mitigate the attack campaign or for forensic analysis. 

\subsection{Selection of Optimal Data Source}
\label{sel-opt-data-src}
As outlined in Section \ref{apt-det-literature}, we utilize data from various sources towards APT detection instead of limiting ourselves to a specific or proprietary data source. However, not all of those data sources are required for detection of each APT stage. Some of the data sources might be redundant or have limitations compared to other data sources. Therefore, we analyse our data sources in the context of each invariant APT stage and select the optimal one.

\begin{itemize}
\item \textit{Command-and-Control}: Most rootkits and malware regularly exchange keep-alive packets with a C\&C server to maintain connection throughout the duration of infection which can be captured by network traffic traces. By default, network IDS rules are not configured to detect C\&C server message exchange. Audit-based provenance which captures network socket file read/write operations can also be used to detect the establishment of a connection between the compromised host and public C\&C server. However, due to the issues of pruning spurious dependencies and noise in provenance graphs due to benign activities, we feel that using audit-based provenance to extract C\&C communication might not justify the computational cost incurred and therefore, network traffic traces alone can be sufficient for this purpose.
\item \textit{Discovery}: This stage can be detected through network traffic traces since scanning for other targets results in TCP/UDP packets being sent from the compromised machine to other machines in the target network. This stage can also be detected from alerts generated by an open source network IDS such as Snort or Suricata. However, not all organizations can be expected to deploy network IDSes. Further, slow stealth scans might be able to evade IDSes altogether. In case network IDS alerts are available, we can use them to increase the accuracy of detection.
\item \textit{Lateral Movement}: This stage can be detected using login logs on target machines since during lateral movement, the attacker logs in to a target machine which is connected to the same or different subnet as the compromised machine. IDS alerts would not be helpful in detecting lateral movement since by default, IDS rules are not configured to do so and lateral movements are not that frequent compared to the events that IDSes are expected to detect such as DoS attacks, port scans, SMB probes, etc. Network traffic traces can contain evidence of lateral movement in terms of TCP/UDP packets exchanged between compromised source and target machines and hence, they can be used to increase the accuracy of lateral movement detection.
\item \textit{Fieldbus scanning}: This stage can be detected through network traffic traces since scanning for target control elements over fieldbus protocols results in TCP packets being sent from the edge gateway to connected control elements at specific port numbers. This stage can also be detected from alerts generated by an open source network IDS such as Snort or Suricata configured with customized modules. In case such IDS alerts are available, we can use them to increase the accuracy of detection.
\item \textit{CE communication spoofing}: Since the attacker needs to send appropriately crafted packets over the link between edge gateway and control elements, network traffic traces can be used to detect the attacker's spoofing of communication with control elements.  
\end{itemize}

%

\subsection{APT Attack-Stage Detection}
\label{apt-attack-det-method}


\textbf{Detection of Command-and-control stage}: In this stage, a compromised machine establishes connection with a C\&C server, which is a public server, and communicates with it at regular time intervals. Therefore, multiple packet exchanges of a suspected host with public server IP address indicates communication with C\&C server.  We filter the TCP {[PSH,ACK], [ACK]}/UDP packets sent/received by public server IP addresses (except VPN server) to/from a given host IP address from network traffic traces (in \textit{pcap} format). If the filter yields multiple such packets, we extract the packet arrival timings. Further, since many malware enforce periodic communication between the compromised machine and C\&C server, we test for periodicity in packet arrival timings obtained earlier using the algorithm proposed in \cite{ayush-cose} which is based on discrete-time signal encoding and autocorrelation function. If the packet arrivals are found to be periodic, it can be inferred that the \textit{Command-and-control} stage has been detected. 

Even if the APT attackers use legitimate protocols to tunnel their commands to a C\&C server, our proposed approach can detect the C\&C communication since the it is protocol-invariant. This has been tested during our performance evaluation of \textsf{RAPTOR} in Section \ref{results} where we use C\&C communication over a legitimate network protocol (DNS) during the APT attack emulation. It should be noted that there may be machines in the enterprise tier which talk to public servers, e.g., to retrieve a web page over HTTPS from a web server. The packets exchanged with such legitimate public servers contribute as noise to our extracted timings for periodicity detection. The ACF (auto-correlation function) can detect periodicity reliably in presence of such noise. 

\textbf{Detection of Discovery stage}:
The network traffic traces collected at a host are split into smaller traces of fixed time duration. These traces are assumed to belong to either of two classes: \textit{normal} or \textit{scanning}. A \textit{normal} trace is one which does not consist of network scanning packets while a \textit{scanning} trace is one that does. Each trace is classified using an ML algorithm (e.g., Decision Trees, SVM, etc.). If a trace is classified as belonging to \textit{scanning} class, it can be inferred that the \textit{Discovery} attack stage has been detected. For each trace, the features for ML classification are extracted from TCP/UDP headers only and not the payloads of the respective packets since the network traffic might be encrypted. The ML features selected for detection of scanning traffic are shown in Table \ref{ml-features}.

\textit{Motivation behind feature selection}: The intuition behind selecting the first two features is that port scanning tools used by attackers send TCP/UDP requests to multiple IP addresses to find out open ports and the services running on them. The third set of features was selected because during port scan targeting an IP address, many of the ports to which TCP connection requests are sent are not open and no response/acknowledgement is sent back, so the TCP connections formed remain half-open. The fourth feature seeks to exploit the fact that once TCP connection is formed with an IP address at a certain port number, port scanning tools exchange data only for a short time and the connection is subsequently reset. Compared to packet length in normal TCP connections, port scanning packet lengths are generally shorter making the fifth set of features useful for classification. Finally, the sixth feature set targets the short time intervals between the transmission of port scanning packets as compared to normal packets.

\textbf{Detection of Lateral movement stage}: According to our approach, the authentication logs on network hosts are used to detect this stage. We look for logins that satisfy the following suspicious property: the machine from which a user is initiating the login to another machine is a part of other detected APT attack stages that usually precede lateral movement, e.g., command-and-control and discovery. For example, if \textit{user1} logs in to \textit{machine-B} from \textit{machine-A}, \textit{machine-A} is in a different subnet as \textit{machine-B} and \textit{machine-A} has been identified as part of \textit{Discovery} stage detected earlier, we can conclude that lateral movement has been detected from \textit{machine-A} to \textit{machine-B}.

\textbf{Detection of Fieldbus scanning}: Similar to the detection of \textit{Discovery} stage, the network traffic traces collected at edge gateway network interfaces connected to control elements are split into smaller traces of fixed time duration. These traces are assumed to belong to either of two classes: \textit{normal} or \textit{fieldbus scanning}. A \textit{normal} trace is one which does not consist of Fieldbus scanning packets while a \textit{fieldbus scanning} trace is one that does. Each trace is classified using an ML algorithm. If a trace is classified as belonging to \textit{fieldbus scanning} class, it can be inferred that the \textit{Fieldbus scanning} attack stage has been detected. For each trace, the features for ML classification are extracted from TCP headers only and not the payloads of the respective packets since the network traffic might be encrypted. The ML features selected for detection of Fieldbus scanning traffic are shown in Table \ref{ml-features}.

\textit{Motivation behind feature selection}: The intuition behind selecting the first three features is that typically during fielbus scanning, attackers attempt to set up a TCP connection with a fieldbus device (PLC/RTU) which is unsuccessful for the first few times. Once the connection is set up, the fieldbus scanner requests device enumeration data and finally, the connection is closed. If the scanner is also trying to enumerate fieldbus slaves, it iterates through the list of slave IDs sequentially. Since slaves are not present at all SIDs, the TCP connection may be reset only for the fieldbus scanner to set up a new connection. Again, compared to packet length in normal TCP connections, fieldbus scanning packet lengths are generally shorter making the fourth set of features useful for classification. Finally, the fifth feature set targets the short time intervals between the transmission of fieldbus scanning packets as compared to normal packets.

\textbf{Detection of CE communication spoofing}: If an attacker is trying to spoof communications with a control element using one of the standard industrial automation (IA) protocols (e.g., IEC 61850, IEC 61131-3), there would either be more than one TCP connections from the edge gateway to the control element at the destination port specific to that IA protocol, or the original TCP connection would be terminated by the attacker leaving only the attacker's TCP connection active. Therefore, if there are more than one TCP connections from the edge gateway to the control element at the destination port specific to that IA protocol or the original TCP connection has been terminated, it can be inferred that the \textit{CE communication spoofing} stage has been detected.
%

\begin{table}[h]
	\centering
  \begin{adjustbox}{width=\columnwidth}
    \begin{tabular}{ | l | p{.6\columnwidth} | }
    \hline
    \textbf{Attack stage} & \textbf{ML features} \\ \hline
    Discovery & \textit{Number of unique TCP SYN/UDP destination IP addresses, Number of unique TCP SYN/UDP destination ports per destination IP address (maximum, minimum, mean), Number of half-open TCP connections, Number of TCP RESET packets, Packet length in bytes (maximum, minimum, mean), Packet inter-arrival time in seconds (maximum, minimum, mean)} \\ \hline
 	Fieldbus scanning & \textit{Number of TCP 3-way handshakes with a destination IP address (maximum, minimum, mean), Number of TCP RESET packets, Number of TCP FIN packets, Packet length in bytes (maximum, minimum, mean), Packet inter-arrival time in seconds (maximum, minimum, mean)} \\ \hline
    \end{tabular}
  \end{adjustbox}
    \caption{ML features selected for detection for APT attack stages}
    \label{ml-features}
\end{table} 

\subsection{Detection Using Multiple Data Sources}
As explained in Section \ref{sel-opt-data-src}, there can be more than one data sources which can be used to detect each APT attack stage. For an attack stage, $a \in$ \{Command-and-control, Discovery, Lateral movement, Fieldbus scanning, CE communication spoofing\}, we define the aggregate detection score as:
\begin{equation}
d_a = \frac{w_{opt} d_{opt}  + \sum_{i} w_{ia} d_{ia}}{w_{opt} + \sum_{i} w_{ia}},
\end{equation}
where $w_{opt}$ is the weight assigned to optimal data source for detection of attack stage $a$, $d_{opt}$ is the detection score given by optimal data source defined as:
\begin{equation}
d_{opt} = 
	\begin{cases}
 		1, & \text{if attack stage $a$ has been detected using}  \\
 		& \text{\hspace{100pt} optimal data source}  \\
		0, & \text{otherwise}
	\end{cases}
\end{equation}
, $w_{ia}$ is the weight assigned to the $i^{th}$ secondary data source for detection of attack stage $a$ (as identified in Section \ref{sel-opt-data-src}), 
and the per-data source detection score, $d_{ia}$ is defined as:
\begin{equation}
d_{ia} = 
	\begin{cases}
 		1, & \text{if attack stage $a$ has been detected using}  \\
 		& \text{\hspace{100pt} $i^{th}$ data source}  \\
		0, & \text{otherwise}
	\end{cases}
\end{equation}

Here, the following condition should always hold: $w_{opt} > w_{ia}$. If the aggregate detection score, $d_a$ is greater than a pre-defined threshold, $\tau$, then the attack stage $a$ is considered as detected, otherwise not. 

\subsection{Correlation of APT Attack Stages}
\label{apt-attack-corr-method}
There are three conditions which need to be satisfied for an attack stage A to be followed by an attack stage B:
\begin{itemize}
	\item The \textit{source IP address} for stage A should match with the \textit{source IP address} for stage B.
	\item When stage A involves movement of the attacker from one machine to another (e.g., Lateral movement), then the \textit{destination IP address} for stage A should match with the \textit{source IP address} for stage B.
	\item The time stamp for stage A should fall earlier than the time stamp for stage B.
\end{itemize}

The APT attack stage detection \& correlation (ASDC) engine first checks if the initial stage of \textit{Command-and-control} in the proposed IASM can be detected at any of the Internet-facing  hosts in enterprise tier. If the detection is successful, ASDC engine checks for the \textit{Discovery} stage at the same host where \textit{Command-and-control} stage was detected. If the \textit{Discovery} stage is detected, ASDC engine looks for signs of the \textit{Lateral movement} stage at the same host.
If the \textit{Lateral movement} stage is detected, ASDC engine proceeds to check for the \textit{Discovery} stage again followed by the \textit{Lateral movement} stage as outlined above at the host accessed after lateral movement. If the host accessed after lateral movement is the edge gateway, ASDC engine starts looking for the \textit{Fieldbus scanning} stage at the edge gateway and if it is detected, ASDC engine checks if \textit{CE communication spoofing} stage can also be detected. The complete attack stage detection and correlation algorithm proposed above is shown in Algorithm \ref{B1}.

\begin{algorithm}[h]
	\centering	
	\caption{\scriptsize{Detect\_Correlate\_Attack\_Stages (list\_host\_IP\_add)}}
	\label{B1}
	\begin{algorithmic}[1]
		\State \textbf{INPUT}: list\_host\_IP\_add (List of IP addresses of Internet-facing hosts)
		\State det\_status $=$  APT\_DET\_START  
		\For {\textit{host\_ip} $\in$ list\_host\_IP\_add}
			\If {\textsf{Check\_C\&C\_stage} (host\_ip)[1] $=$ TRUE}
				\State src\_host\_ip $=$ host\_ip
			\EndIf
		\EndFor
		\If {\textsf{Check\_Discovery\_stage} (src\_host\_ip)[1] $=$ TRUE \&\& \textsf{Check\_Discovery\_stage} (src\_host\_ip)[2] $>$ \textsf{Check\_C\&C\_stage} (src\_host\_ip)[2]}
			\For {tgt\_host\_ip $\in$ \textsf{Check\_Discovery\_stage} (src\_host\_ip)[4]}
				\If {\textsf{Check\_Lateral\_Movement\_stage} (src\_host\_ip, tgt\_host\_ip)[1] $=$ TRUE \&\& \textsf{Check\_Lateral\_Movement\_stage} (src\_host\_ip, tgt\_host\_ip)[2] $>$ \textsf{Check\_Discovery\_stage} (src\_host\_ip)[2]}
					\If {tgt\_host\_ip $=$ edge\_gw\_IP}
						\If {\textsf{Check\_Fieldbus\_scan\_stage} ()[1] $=$ TRUE \&\& \textsf{Check\_Fieldbus\_scan\_stage} ()[2] $>$ \textsf{Check\_Lateral\_Movement\_stage} (src\_host\_ip, tgt\_host\_ip)[2]}
							\If {\textsf{Check\_CE\_comm\_stage} ()[1] $=$ TRUE \&\& \textsf{Check\_CE\_comm\_stage} ()[2] $>$ \textsf{Check\_Fieldbus\_scan\_stage} ()[2]}
								\State det\_status $=$  APT\_DET\_STOP
								\State return (det\_status)
							\EndIf
						\EndIf
					\EndIf
				\EndIf
			\EndFor	
		\EndIf
		
		\State \textbf{Function} \textsf{Check\_C\&C\_stage} (host\_ip), \textit{Returns} \{bool\_val, time\_det, C\&C\_server\_IP\}
		\State \textbf{Function} \textsf{Check\_Discovery\_stage} (host\_ip), \textit{Returns} \{bool\_val, time\_det, scan\_type, list\_target\_host\_IPs\}
		\State \textbf{Function} \textsf{Check\_Lateral\_Movement\_stage} (src\_host\_ip, dst\_host\_ip), \textit{Returns} \{bool\_val, time\_det\}
		\State \textbf{Function} \textsf{Check\_Fieldbus\_scan\_stage} (), \textit{Returns} \{bool\_val, time\_det\}
		\State \textbf{Function} \textsf{Check\_CE\_comm\_stage} (), \textit{Returns} \{bool\_val, time\_det\}
	\end{algorithmic}
\end{algorithm}

\textit{Handling false positives/negatives in ML classification}: \textsf{RAPTOR} is designed to handle false positives/negatives in ML-based detection. For example, let us assume that there is a false positive, i.e., a packet trace is classified as \textit{scanning} though it is \textit{normal} and the \textit{Discovery} stage is marked as detected. The subsequent attack stages in IASM would not be detected by the ASDC engine and therefore, \textsf{RAPTOR} would know that there was a false positive in detection of an earlier attack stage. It is also possible that there is a false negative, i.e., a packet trace is classified as \textit{normal} though it is \textit{scanning}, and therefore, the ASDC engine would not invoke detection of subsequent attack stages in IASM. We propose to handle both false positives and negatives by taking the mode of classification results for a packet trace for a sufficiently large number of iterations.

\subsection{APT Campaign Graph}
\label{apt-campaign-graph}
As the ASDC engine proceeds to detect various stages of an APT campaign, it uses the detected stages and their attributes to construct the APT campaign graph. It is a directed graph, $G(V,E)$ where each node, $v_i \in V$ $\forall i \in \{1,2,...,N_v\}$, where $N_v$ is the total number of nodes in the graph, corresponds to a machine (denoted by its IP address) which is a part of one of the detected attack stages. An edge, $e_j \in E$ $\forall j \in \{1,2,...,N_e\}$, where $N_e$ is the total number of edges in the graph, is extended from node $v_i$ to another node $v_k$ in the graph if there is a connection from the machine corresponding to node $v_i$ to the machine corresponding to node $v_k$ during one or more of the APT attack stages. Each edge has an attribute $\{s_1, s_2,...\}$ where $s_l$ $\forall l \in \{1,2,...\}$ is an attack stage which enables the connection between the two machines corresponding to the nodes at either end of the edge. 


\section{Evaluation}

\subsection{IIoT Testbed}
\label{testbed}
\begin{figure}[h]
	\centering
	\includegraphics[scale=0.15]{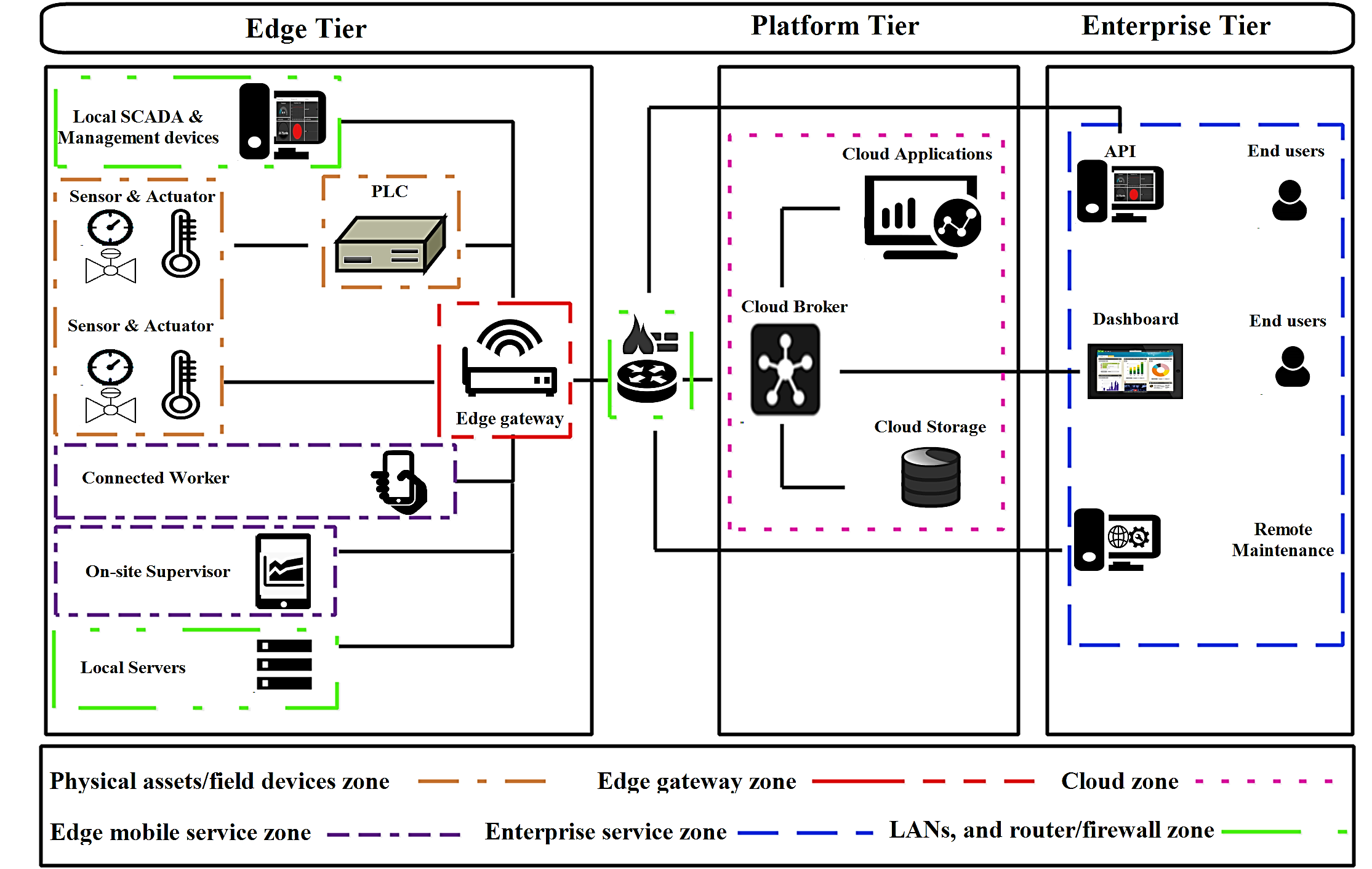}
	\caption{IIoT Testbed Architecture \cite{brown-iiotbed}}
	\label{brown-iiotbed-arch}
\end{figure}

To generate a realistic IIoT APT dataset which can be used to evaluate \textsf{RAPTOR}'s performance, we built an IIoT testbed modelled after Brown-IIoTbed \cite{brown-iiotbed} whose architecture is reproduced in Fig. \ref{brown-iiotbed-arch}. We decided not to use the existing X-IIoTID intrusion dataset \cite{x-iiotid} which also uses the Brown-IIoTbed testbed and covers various attack scenarios and attacks related to IIoT connectivity protocols. This is because X-IIoTID provides pre-decided ML features extracted from network flows only and does not include raw traffic traces from which additional features can be extracted. Further, the dataset provides host logs and IDS alert logs collected at the edge gateway only and does not include APT campaigns.

The implementation of IIoT testbeds is still in its early stages, with most existing implementations [4] being special projects and publicly unavailable. Brown-IIoTbed is designed based on the IIC (Industrial Internet Consortium)'s IIRA (Industrial Internet Reference Architecture) model and consists of three tiers- \textit{edge}, \textit{platform} and \textit{enterprise}. It supports a number of real-world IIoT functionalities such as  e-mail notifications to plant workers regarding important OT events, web-based SCADA interface (viewing real-time sensor values and trends, actuator status change notifications, tuning of PLC parameters), remote maintenance of edge gateway, query to edge data historian, etc. The testbed also supports a number of real-world IIoT protocols such as CoAP, MQTT, and Modbus. 

\subsection{Experimental Methodology and Performance}
\label{results}
We collect data from our IIoT testbed under normal operation as well during APT attack stages. Data is collected from our testbed in the form of network traffic traces from hosts, audit-based provenance at each x86-based host, host logs (login records, authentication logs, syslog) and alerts from Snort network IDS. However, we use only the optimal data sources identified in Section \ref{sel-opt-data-src} towards the final attack detection.


\textbf{Command-and-control stage}: To emulate this stage, we use open-source tools such as \textit{dnscat2} to create a communication channel between a C\&C server and a compromised machine using tunneling over DNS protocol which is one of the most common C\&C communication protocols used by attackers since most firewalls do not block it. We run a public \textit{dnscat2} C\&C server and the client on a Windows 10 Pro and a Ubuntu 20.04 machine in our testbed. The packet traces generated on those client machines are collected using \textit{tcpdump} in sets of 1 minute duration. A total of 1000 packet traces were collected for each type of host and fed as input to the detection phase of \textit{Algorithm 2} proposed in \cite{ayush-cose} for detection of C\&C communication.

\textit{Results}: The performance is evaluated in terms of detection rate (DR) and missed-detection rate (MDR). \textit{Detection Rate} is the fraction of the total number of packet traces which have been correctly detected as containing C\&C traffic, and \textit{Missed-detection Rate} is the fraction of the total number of packet traces which have been incorrectly detected as not containing C\&C traffic. Using the parameter values (given in Table \ref{algo2params}), the detection performance for both the scenarios specified above is shown in Table \ref{cncdet-scores}. It can be seen that the algorithm gives a DR of $1.0$ and a MDR of $0.0$. 

\begin{table}[h]
	\centering
    \begin{tabular}{ | l | l |}
    \hline
    Traffic sampling frequency & 0.1 \\ \hline
    Min. autocorrelation peak height & 0.7$\times$(Max. peak height) \\ \hline
    Inter-peak gap variance threshold & 0.01 \\ \hline
    \end{tabular}
    \caption{Parameter Values for C\&C Communication Periodicity Detection}
    \label{algo2params}
\end{table}

\begin{table}[]
	\centering
	\begin{tabular}{|l|l|l|l|}
	\hline
	\textbf{DATASET} & \textbf{METHOD} & \textbf{DR} & \textbf{MDR} \\ \hline
	{IIoT Testbed Ubuntu host} & Algorithm 2 \cite{ayush-cose} & 1.0 & 0.0 \\ \hline
	{IIoT Testbed Windows host} & Algorithm 2 \cite{ayush-cose} & 1.0 & 0.0 \\ \hline
	\end{tabular}
	\caption{C\&C Stage Detection Performance}
	\label{cncdet-scores}
\end{table}


\textbf{Discovery stage}: To emulate this stage, we use open-source network scanning tools such as \textit{nmap} which is either used directly or is the inspiration for customized port scanners used by most APT groups. We run the default \textit{nmap} SYN scans on a Windows 10 Pro and a Ubuntu 20.04 machine in our testbed to enumerate connected hosts, their OS versions and the services running on them. \textit{Nmap} is run in both \textit{normal} mode as well as \textit{sneaky} or as we call it, slow mode, with the latter mode targeted at evading IDSes \cite{nmap-speeds}. The packet traces thus generated on those machines are collected using \textit{tcpdump} in sets of 1 minute duration. In real-world APT campaigns, attackers may slow down network scanning to evade detection by IDS and therefore, we may need to increase the duration of packet captures. A total of 1000 packet traces are collected from both the Windows and Ubuntu machines under normal operation and further 1000 packet traces are collected during the network scanning operation. The packet traces are used to extract features mentioned in Section \ref{apt-attack-det-method} and appropriate class labels ('normal' or 'scanning') are assigned to them. The extracted features vectors are further processed (handling of missing values, scaling) and randomly divided into \textit{training} and \textit{test} datasets using an 80:20 split. Using $\chi^2$ test statistic, we select the best features (test statistic value above a pre-selected threshold) out of the existing ones. The final feature vectors thus obtained are used to train Support Vector Machine (SVM) and Random Forest (RForest) models. The trained ML models are then used to predict class labels for the test dataset and finally, the detection performance of the models is evaluated. We use a 10-fold cross validation approach to tune the hyper-parameters of the ML classifiers for achieving the highest possible CV scores. The cross validation is based on training data only without using any information from the test dataset.

\textbf{Fieldbus scanning stage}: To emulate this stage, we run \textit{nmap} with \textsf{modbus-discover} script on the edge-gateway for enumerating Modbus slave IDs and collecting details about the slave devices. We run the \textsf{modbus-discover} script in both 'aggressive' and 'non-aggressive' modes, where the former mode refers to finding all slave IDs and the latter mode refers to finding just the first slave ID. Though Modbus is one of the common protocols used for communication with PLCs/RTUs, there are other protocols as well which are used in the industry, e.g., DNP3 (Distributed Network Protocol), Profibus/Profinet, CAN (Controller Area Network). Therefore, in a separate experiment, we connect a Seimens S7-1200 PLC to the edge gateway network and run \textit{nmap} with \textsf{s7-info} script on the gateway for enumerating Seimens S7 PLC devices and collecting their device information. The steps for packet trace collection under normal and fieldbus scanning operations, feature vector extraction, processing and selection, ML model training and performance evaluation remain similar to the ones outlined for \textit{Discovery} stage above.


\begin{table}[]
	\begin{tabular}{|l|l|l|l|}
	\hline
	\textbf{DATASET} & \textbf{MODEL} & \textbf{PR} & \textbf{RC} \\ \hline
	\multirow{2}{*}{\thead{IIoT testbed Ubuntu host\\ (Discovery-normal)}} & Rforest & 0.996 & 1.0 \\
	 &  SVM & 1.0 & 1.0 \\ \hline
	 	\multirow{2}{*}{\thead{IIoT testbed Ubuntu host\\ (Discovery- slow)}} & Rforest & 0.991 & 1.0 \\
	 &  SVM & 0.978 & 0.974 \\ \hline
	 	\multirow{2}{*}{\thead{IIoT testbed Windows host\\ (Discovery-normal)}} & Rforest & 0.996 & 1.0 \\
	 &  SVM & 0.979 & 1.0 \\ \hline
	 	\multirow{2}{*}{\thead{IIoT testbed Windows host\\ (Discovery- slow)}} & Rforest & 1.0 & 1.0 \\
	 &  SVM & 0.912 & 0.978 \\ \hline
	 \multirow{2}{*}{\thead{IIoT testbed\\ (Fieldbus (Modbus) Scanning- agg.)}} & Rforest & 1.0 & 0.992 \\
	 &  SVM & 1.0 & 0.996 \\ \hline
	 	 \multirow{2}{*}{\thead{IIoT testbed\\ (Fieldbus (Modbus) Scanning- non-agg.)}} & Rforest & 0.996 & 1.0 \\
	 &  SVM & 0.996 & 0.983 \\ \hline
	 	 \multirow{2}{*}{\thead{IIoT testbed\\ (Fieldbus (Profibus) Scanning)}} & Rforest & 0.992 & 1.0 \\
	 &  SVM & 0.988 & 0.996 \\ \hline
	\end{tabular}
	\caption{Raptor's ML performance for detection of Discovery and Fieldbus scanning stages}
	\label{class-scores}
\end{table}


\textit{Results}: The performance of ML classifiers is typically evaluated in terms of precision (PR) and recall (RC) scores.
\textit{Precision} is the ratio $\frac{TP}{TP+FP}$, where $TP$ is the number of true positives and $FP$ is the number of false positives. It represents the ability of a classifier to avoid labeling samples that are negative as positive. \textit{Recall} is the ratio $\frac{TP}{TP+FN}$, where $TP$ is the number of true positives and $FN$ is the number of false negatives. It represents the ability of a classifier to avoid labeling samples that are positive as negative. Using the tuned hyper-parameters' values, the average classification precision (PR) and recall (RC) scores obtained for the final classifiers over 10 runs are shown in Table \ref{class-scores}. It can be observed that for the detection of \textit{Discovery} stage on Ubuntu as well as Windows hosts using normal and slow scan speeds,  Random Forest performs slightly better than SVM. In general, both the ML classifiers perform better with normal scanning speed compared to slow scanning speed which is expected since within the trace duration (1 min), more number of network scanning packets would be captured during normal versus slow speed scanning.
For the detection of \textit{Fieldbus scanning} stage using Modbus protocol in 'aggressive' and 'non-aggressive' modes, Random Forest performs almost equally as SVM in terms of precision but SVM performs quite poorly compared to Random Forest in terms of recall for 'non-aggressive' mode. For the detection of \textit{Fieldbus scanning} stage using Profibus protocol, Random Forest performs slightly better than SVM in terms of both precision and recall. Based on the performance results obtained above, it would be preferable to select Random Forest classifier for detection of \textit{Discovery} and \textit{Fieldbus scanning} stages in \textsf{RAPTOR}'s implementation since SVM's performance degrades significantly at slow network scanning speed and for 'non-aggressive' fieldbus scanning mode.

\textbf{APT Campaign Graph Construction}: To emulate an APT attack on our testbed for construction of APT campaign graphs, we develop three attack campaigns from an APT group’s perspective, i.e., their background, motivation for attack, steps taken for attack and final attack objective. The TTPs used in our campaigns are close to the ones used in real-world APT attacks on IIoT environments such as those mentioned in Section \ref{intro}. For reasons of space, we present \textsf{RAPTOR}'s evaluation with only one of the APT attack campaigns here. 
The rest of the campaigns are presented in Appendix Sections \ref{apt-campaign2} and \ref{apt-campaign3}. 
Steps 1-2 (Initial Access), step 4 (Command-and-control), step 6 (Discovery) and step 10 (Credential Access, Lateral Movement) of the attack campaign are based on the 2014 German steel mill and 2015/2016 Ukraine power grid attacks. Step 16 (Fieldbus scanning, CE communication spoofing) of the attack campaign is based on the 2016 Ukraine power grid attack and the 2017 Saudi petrochemical plant attack. Steps 18-19 (Impact) of the attack campaign are based on the 2015 Ukraine power grid attack. We run the attack campaign on our IIoT testbed over the course of a few hours and collect the data generated from testbed hosts. Since real-world APT campaigns can stretch over months and it is not possible to emulate them on our testbed, we assume that our APT campaigns are executed in an accelerated timeframe and therefore our performance evaluation of \textsf{RAPTOR} holds. The complete APT attack campaign used for evaluating \textsf{RAPTOR} is as follows.


\textit{APT Background}: Attackers belong to a nation-state (or APT) group which has been tasked with targeting a prominent state-owned steel manufacturing plant. The APT group plans to steal ICS related data which can be used to understand the ICS design and components which can further be used to plan for later attacks.

\textit{APT Goals}: To steal sensitive OT data (e.g., blast furnace temperature sensor measurements, PLC configuration, credentials). OT data such as blast furnace temperature readings are sensitive because they can be used to learn the normal temperature range and temporal trends. Attackers can use this information to modify the settings of the furnace temperature controller to damage the furnace. The temperature readings can also be used to infer the furnace design. 

\textit{APT Campaign Steps}:
\begin{enumerate}
	\item The attacker sends a spear phishing email (including a malicious VPN portal web link) to one of the steel plant employees posing as legitimate company email and obtains their VPN login credentials.
	\item It uses the employee’s company email address and phished credentials to remotely login to the maintenance machine connected to enterprise network (password re-use) through the VPN service.
	\item The attacker changes the employee’s VPN account password for persistence.
	\item The compromised machine connects to an external C\&C server through DNS tunnelling and forwards a shell to the attacker.
	\item Attacker installs a malware on the compromised machine which exploits software vulnerabilities to gain root access.
	\item The attacker controlling the compromised machine scans its local network and finds other hosts (firewall, MQTT server, external API machine) as well as the services running on them.
	\item Attacker tries to find CVE vulnerabilities corresponding to the services running on other hosts but can not exploit them successfully.
	\item It goes through the shell command history on compromised machine and finds previous SSH connection attempts to the edge gateway containing \textit{username} and \textit{hostname} details.
	\item It tries to determine the SSH login password for the edge gateway as follows:
	\begin{enumerate}
		\item Accesses the shadow password file on compromised machine (using root access obtained earlier) which stores password hashes and corresponding hashing algorithms used.
		\item Tries to crack the password hashes to obtain corresponding plaintext passwords.
	\end{enumerate}
	\item The attacker attempts to log in to the edge gateway by using one plaintext password at a time and is successful. It explores the files, folders (hidden and non-hidden) and the processes running on edge gateway.
	\item It finds a web server, a CoAP server and Node-red application running on edge gateway.
	\item The attacker tries to exploit CoAP related vulnerabilities but is unsuccessful. It remotely executes a script from the compromised maintenance machine to dump the CoAP resources.
	\item It executes a fake CoAP client code on the compromised maintenance machine to receive measurements from sensors directly connected to edge gateway. 
	\item It scans devices connected to the edge gateway’s Wi-Fi hotspot network and finds a host running DNP service (PLC master).
	\item Attacker downloads a script from C\&C server and copies it remotely to edge gateway.
	\item It extracts PLC configuration data (e.g., hardware, firmware, manufacturer, serial number, slave IDs) by running the script on edge gateway.
	\item The attacker compresses and encrypts all the targeted data collected in previous steps (e.g., PLC configuration, sensor measurements, login credentials) and exfiltrates it through the C\&C channel.
\end{enumerate}

%
%

\textit{Graph Analysis}: We assigned a weight, $w_{opt}=0.5$ to the optimal data source for detecting an attack stage and a weight of of $w_{ia}=0.25$ to the secondary data sources. The threshold for detection is selected as $\tau=0.5$. The APT campaign graph generated by \textsf{RAPTOR} for the attack campaign outlined above is shown in Fig. \ref{apt-campaign-graph}. The graph captures broad details of the APT campaign including the IP addresses of the machines affected and the tactics used during the campaign which is quite useful for cybersecurity analysts. This shows that our proposed attack stage detection and correlation algorithm in Section \ref{apt-attack-corr-method} works as intended. However, the campaign graph does not capture all the tactics employed by the APT attackers since our focus is on detecting invariant APT tactics/stages only as explained in Section \ref{iasm}. Further, the campaign graph does not contain any details on the specific techniques employed by the APT attackers since our focus is not on detecting the individual techniques used for each tactic. The APT campaign graph can serve as a starting point for cybersecurity analysts to fill in the missing tactics based on the APT attack frameworks for ICS, further investigation and mitigation. 

\begin{figure}[h]
\centering
\includegraphics[scale=0.3]{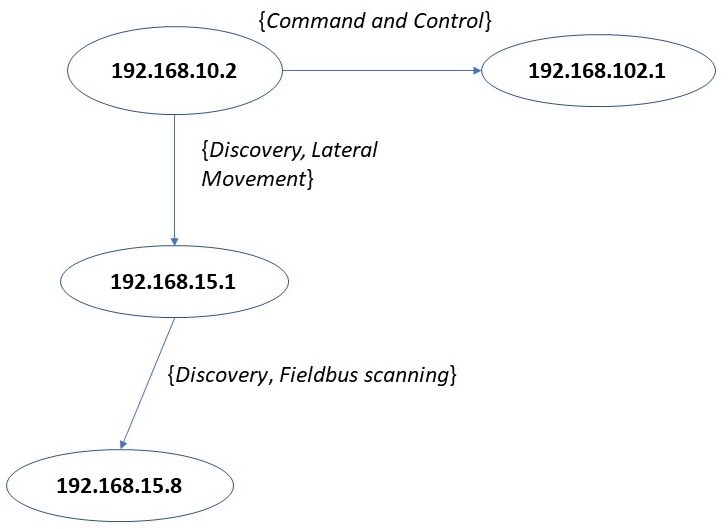}
\caption{APT campaign graph generated for \textit{Campaign 1}}
\label{apt-campaign-graph1}
\end{figure}  

\section{Comparison with State-of-the-art}
We are unable to conduct a performance comparison of \textsf{RAPTOR} with \cite{kcsm} as they do not provide any source code for their proposed multi-stage attack detection algorithm. Further, the dataset used for performance evaluation in \cite{kcsm} which consist of a synthetic APT campaign injected into the CSE-IDC-2018 intrusion detection dataset \cite{cseidc2018} has not been publicly released. HOLMES \cite{holmes} and CONAN \cite{conan} do not provide source codes for their proposed APT detection system as well though both use the DARPA Transparent Computing ((TC) Engagement dataset \cite{darpa-tc} for performance evaluation which has been released publicly. The DARPA TC dataset contains data from a red team deploying APT-style TTPs on a target system consisting of multiple interconnected hosts running different OSes and having exploitable CVE vulnerabilities. However, the DARPA TC dataset suffers from following limitations which reduce its applicability for \textsf{RAPTOR}'s performance evaluation:
\begin{itemize}
	\item The TC network setup is simple and does not emulate real-world enterprise/IIoT networks.
	\item The dataset provides only \textit{json} files but no raw \textit{pcap} files for us to extract network traffic-based features for a meaningful comparison with \textsf{RAPTOR}'s performance on our IIoT testbed dataset\footnote{There is an active unresolved issue with the DARPA TC dataset. While loading data from the compressed *.bin.1.gz files, the code gets stuck at streaming records.}.
	\item None of the TTPs used in TC dataset are IIoT-specific.
\end{itemize}

The mechanisms used for detection of \textit{Command-and-control} and \textit{Discovery} stages in an APT campaign by \textsf{RAPTOR} are similar to those used in \cite{ayush-cose} for detection of IoT botnets since the botnets use those two attack tactics for sending commands to bots and botnet propagation. Thus, the limitations of the IoT botnet detection approach and the ways to address them apply to \textsf{RAPTOR} as well.

\section{Conclusion} 
We have proposed \textsf{RAPTOR}, an APT detection system targeted at IIoT environments. It detects and correlates attack stages derived from an APT Attack Invariant State Machine using optimal data sources selected for each stage. The correlated attack stages are utilized to generate a compact, high-level APT Campaign Graph which can be used by cybersecurity analysts to track the progress of the APT campaign and deploy appropriate mitigation measures. 
A performance evaluation of \textsf{RAPTOR} shows that it can detect APT campaigns modelled after real-world attacks with high precision and low false positive/negative rates.			 


\bibliographystyle{ieeetran}
\begingroup
\raggedright
\bibliography{raptor}
\endgroup

\appendices

\section{APT campaign 2}
\label{apt-campaign2}

\textit{APT Background}: Attackers belong to a nation-state (or APT) group which has been tasked with targeting a prominent state-owned steel manufacturing plant. The APT group plans to disrupt the steel production and thereby affect other industries dependent on steel and exports.

\textit{APT Goals}: To shut the blast furnace down by controlling the furnace relays (LEDs in our testbed). This may damage the plant operations temporarily or permanently.

\textit{APT Campaign Steps}:
\begin{enumerate}
	\item An insider recruited by the APT group installs malware on the maintenance machine through a USB stick. The malware exploits software vulnerabilities on the machine to gain root access.
	\item The compromised machine connects to an external C\&C server through DNS tunnelling and forwards a remote display to the attacker.
	\item Attacker installs a malware on the compromised machine which exploits software vulnerabilities to gain root access.
	\item The attacker controlling the compromised machine scans its local network and finds other hosts (firewall, MQTT server, external API machine) as well as the services running on them.
	\item Attacker tries to find CVE vulnerabilities corresponding to the services running on other hosts but can not exploit them successfully.
	\item It accesses the shadow password file on compromised machine (using root access obtained earlier) which stores password hashes and corresponding hashing algorithms used.
	\item Attacker successfully opens an RDP (Remote Desktop Protocol) session to the external API machine using one of the stolen password hashes.
	\item It accesses the SCADA/HMI web interface on the external API machine and turns off the LEDs directly connected to edge gateway.
\end{enumerate}

\begin{figure}[h]
\centering
\includegraphics[scale=0.4]{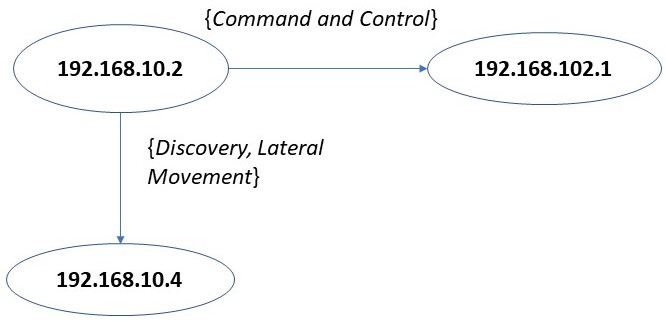}
\caption{APT campaign graph generated for \textit{Campaign 2}}
\label{apt-campaign-graph2}
\end{figure} 

\section{APT campaign 3}
\label{apt-campaign3}

\textit{APT Background}: Attackers belong to a nation-state (or APT) group which has been tasked with targeting a prominent state-owned steel manufacturing plant. The APT group plans to disrupt the steel production and thereby affect other industries dependent on steel and exports.

\textit{APT Goals}: To damage the plant equipment by tampering with the operation of safety controllers which prevent the blast furnace from entering an unsafe state. The safety controllers may be reprogrammed to allow the blast furnace to enter a dangerous state without any corrective action leading to physical damage to the plant and even loss of human lives.

\textit{APT Campaign Steps}:
\begin{enumerate}
	\item The attacker sends a spear phishing email (including a malicious VPN portal web link) to one of the steel plant employees posing as legitimate company email and obtains their VPN login credentials.
	\item It uses the employee’s company email address and phished credentials to remotely login to the maintenance machine connected to enterprise network (password re-use) through the VPN service.
	\item The attacker changes the employee’s VPN account password for persistence.
	\item The compromised machine connects to an external C\&C server through DNS tunnelling and forwards a shell to the attacker.
	\item Attacker installs a malware on the compromised machine which exploits software vulnerabilities to gain root access.
	\item The attacker controlling the compromised machine scans its local network and finds other hosts (firewall, MQTT server, external API machine) as well as the services running on them.
	\item Attacker tries to find CVE vulnerabilities corresponding to the services running on other hosts but can not exploit them successfully.
	\item It goes through the shell command history on compromised machine and finds previous SSH connection attempts to the edge gateway containing \textit{username} and \textit{hostname} details.
	\item It hijacks any future SSH session between the compromised machine (started by an employee performing remote maintenance) and the edge gateway.
	\item The attacker explores the files, folders (hidden and non-hidden) and the processes running on edge gateway.
	\item It finds a web server, a CoAP server and Node-red application running on edge gateway.
	\item The attacker scans devices connected to the edge gateway’s Wi-Fi hotspot network and finds a host running DNP service (PLC device).
	\item It downloads a payload from C\&C server, copies it remotely to edge gateway and executes it.
	\item The attacker terminates the existing process which is communicating with the PLC device.
	\item It collects more information about the PLC device and enumerates all the slave IDs using the payload commands.
	\item The attacker uses the payload to send a command to the targeted slave to read its current state.
	\item It remotely uploads a new program to the PLC device while it continues to operate. 
\end{enumerate}

\begin{figure}[h]
\centering
\includegraphics[scale=0.4]{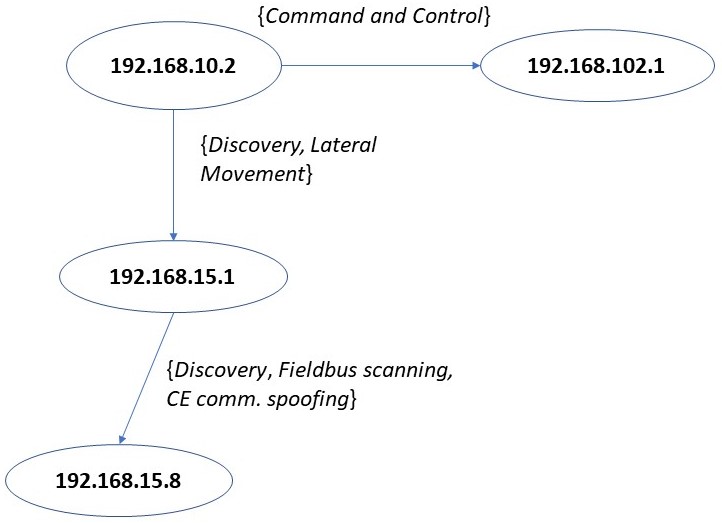}
\caption{APT campaign graph generated for \textit{Campaign 3}}
\label{apt-campaign-graph3}
\end{figure} 

\end{document}